# Designer quantum states of matter created atom-by-atom

Alexander A. Khajetoorians[1,*], Daniel Wegner[1], Alexander F. Otte[2], Ingmar Swart[3,#]

[1.]Scanning Probe Microscopy Department, Institute of Molecules and Materials, Radboud University, Nijmegen, The Netherlands
[2.]Department of Quantum Nanoscience, Kavli Institute of Nanoscience, Delft University of Technology, Delft, The Netherlands.
[3.]Debye Institute for Nanomaterials Science, Utrecht University, Utrecht, The Netherlands
*a.khajetoorians@science.ru.nl
#I.Swart@uu.nl

**With the advances in high resolution and spin-resolved scanning tunneling microscopy as well as atomic-scale manipulation, it has become possible to create and characterize quantum states of matter bottom-up, atom-by-atom. This is largely based on controlling the particle- or wave-like nature of electrons, as well as the interactions between spins, electrons, and orbitals and their interplay with structure and dimensionality. We review the recent advances in creating artificial electronic and spin lattices that lead to various exotic quantum phases of matter, ranging from topological Dirac dispersion to complex magnetic order. We also project future perspectives in non-equilibrium dynamics, prototype technologies, engineered quantum phase transitions and topology, as well as the evolution of complexity from simplicity in this newly developing field.**

**Introduction**

One promising route toward understanding novel quantum states of matter is to engineer many-body states, from the bottom-up, in so-called artificial lattices. 2D artificial lattices can serve as model systems to realize various properties seen in 3D, with the advantage that all the relevant interactions can be controlled, serving as an ideal model system for theoretical innovation[3-5]. The concept behind artificial lattices is building up periodic structures where symmetry and interactions, for example between collections of electrons and/or spins, are precisely tailored[3,6]. Artificial lattices have been realized in vastly different experimental communities, e.g. cold atoms[8], trapped ions[10], superconducting circuits[13], patterned devices[15], optical lattices[16,17], and most recently twisted bilayer graphene[18,19]. Since novel quantum states of matter arise in certain well-defined arrangements of atoms, it is natural to develop platforms that allow the creation and manipulation of quantum matter on the atomic scale with a high degree of tunability, simultaneously enabling characterization with high precision.

The ultimate flexibility is to realize an experiment that combines *in-situ* construction of an electron or spin lattice with a chosen geometry and with atomic-level control of the relevant interactions, while having the simultaneous ability to probe the changes in the electronic and magnetic properties. To this end, atomically precise structures have been created using self-assembly and on-surface chemistry[20-23], where the properties of the final material (which can be the molecular network itself or the surface patterned into an array of quantum confinements) are encoded in the precursor species used. Key strategies are based on utilization of tailored intermolecular interactions (e.g. hydrogen and halogen bonding)[24-28], the formation of metal-organic coordination bonds (e.g. using functional cyano or carboxylate groups)[23,28-31], or on-surface polymerization (e.g. utilizing Ullmann-type reactions)[22,32-35]. An advantage of utilizing self-assembly and self-organization is that large artificial structures can be created quickly. A disadvantage is the reduced flexibility in tuning different structures using the same building blocks, and controlled construction of lateral artificial heterostructures remains difficult. Likewise, incorporation of specific atomically precise defects, as well as the number and spacing of these defects is not possible. Moreover, only few studies thus far focused on tailoring and studying the electronic or spin structure of self-assembled artificial lattices[24,25,27,34,35]. An alternative, especially appealing approach to study quantum matter with ultimate flexibility and precision is artificial lattices derived from patterned atomic impurities, created and probed by scanning tunneling microscopy (STM).[3-5,36]. To date, this concept has been utilized to realize many novel states of matter, ranging from lower-dimensional Bloch states toward Dirac materials such as artificial graphene[3] and the Lieb lattice[4,5], as well as a variety of tailored nanomagnets with designed magnetic order[6,37-40]. This review will hence focus on artificial electronic and spin lattices created by STM-based atomic manipulation. Low-temperature STM has become a powerful tool to both create *and* characterize quantum matter at the atomic scale[41]. Tailoring electronic and magnetic properties with single atoms has a long history[6,38,42,43], stemming from the first quantum corral built in 1993 at IBM Almaden[43] (see info box 2). Since that time, there has been an exhibition of ground-breaking examples, in which manipulated structures have led to new understandings of magnetism, as well as prototype technologies[6,37,38,44,45]. Scanning tunneling spectroscopy (STS) gives access to the density of states, and in combination with

*Box 1*: **The scanning tunneling microscope – an integrated nanolab**

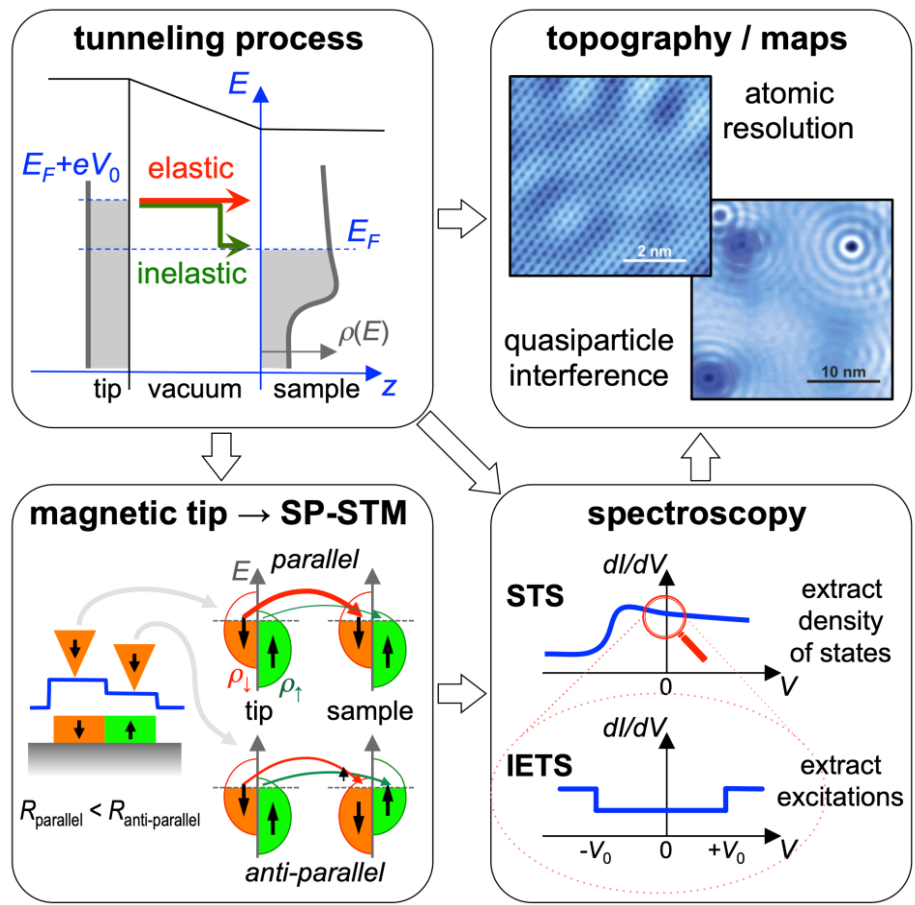

Since its invention and demonstration of atomic resolution by Binnig and Rohrer, the STM has developed into a versatile tool reaching far beyond mere imaging of surfaces, mostly owing to the implementation into cryogenic ultrahigh vacuum environments[51]. Performing STM at low temperatures not only freezes adatom diffusion but also inhibits thermal drift between the STM tip and the sample, permitting to position and stabilize the tip over an atomic site for weeks. This high stability is also the basis for the STM's capability to manipulate single adsorbates (atoms or molecules) or surface vacancies with atomic-scale precision (see Box 2).

The vast success of cryogenic STM is further based on scanning tunneling spectroscopy (STS), where the tunnel current $I$ is measured as a function of applied sample bias $V$ under open-feedback conditions [52,53]. The differential conductance $dI/dV$ is, in good approximation, proportional to the sample's electronic local density of states, allowing to probe both occupied ($V < 0$) and unoccupied ($V > 0$) states. Here, the energy resolution is directly proportional to the temperature ($\Delta E \approx 3.5\, k_B T$), i.e. resolutions of 1-2 meV can be reached with conventional LHe bath cryostat systems, while state-of-the-art $^3$He/$^4$He dilution refrigerator-based systems have shown resolutions more than one order of magnitude better [54-60]. Moreover, inelastic electron tunneling spectroscopy (IETS)[61-63] can be applied on the local scale in STM in order to excite molecular vibrations[64], phonons[65,66], spin excitations[46] or magnons[67]. The classic hallmark of IETS is two symmetric steps around $V = 0$ in $dI/dV$ at the resonance energy $E_0$. The inelastic tunneling leads to mode excitation via energy loss and a lower final-state energy of the tunneling electron, hence opening an additional tunneling channel which leads to a sudden change of conductance at $V_0 = \pm E_0/e$.

A final key ingredient expanding the possibilities of STM is tip functionalization. The most influential type of functionalization is the use of a magnetic tip, by either using a bulk magnetic tip, by coating a nonmagnetic tip with a thin (anti-) ferromagnetic film, or by transferring a single magnetic atom onto the tip in an external magnetic field[68]. The spin-polarized STM technique (SP-STM) is based on the fact that the spin of elastically tunneling electrons is conserved[69,70]. If the tip and sample density of states are spin-polarized, this leads to a tunneling magnetoresistance and the tunnel current gives information on the projected local sample magnetization onto the polarization of the tip's last apex atom, hence enabling magnetic contrast with atomic resolution.

magnetic fields, it has been possible to observe topological superconductivity, Landau quantization in novel materials, and excitations of atomic spins with ultimate spatial resolution[46-49]. The advancements in atomic manipulation and automation[42,50], as well as ultra-high resolution and spin-sensitive scanning tunneling spectroscopy down to mK temperatures (see info box 1), has opened up the possibility to create quantum matter with designed properties and probe these states with atomic-scale resolution in and out of equilibrium. This combination provides the advantage of being able to combine fabrication, tailoring of individual interactions, and subsequent characterization in one seamless experiment.

**Methods**

**Platforms for the realization of artificial lattices**

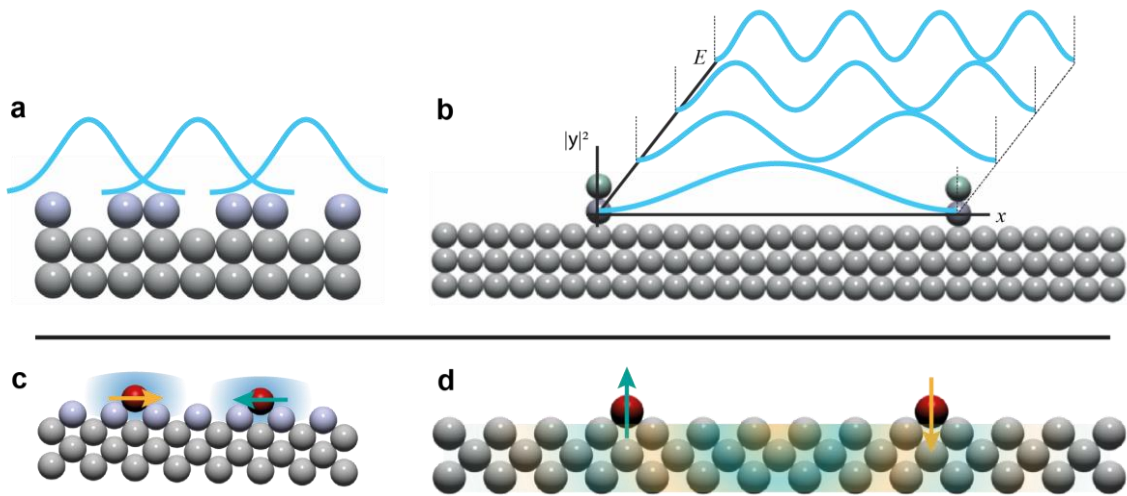

*Fig. 1. Platforms for artificial electronic and spin lattices.* *Two different concepts can be applied based on either localized (left) or delocalized (right) interactions, for both electronic (top) and spin (bottom) lattices. (a) Tight-binding approach utilizing nearest-neighbour overlap of adsorbate orbitals or, as shown here, vacancy states. (b) Nearly free electron approach using 2D electron gases confined and restructured by adsorbates that serve as scattering centres. Bottom panel: spin lattice concepts. (c) Nearest-neighbour direct or superexchange of localized spins. (d) Long-range RKKY-like exchange interaction mediated by delocalized conduction band electrons.*

There are two complementary models to describe the electronic structure of artificial lattices, namely the tight-binding and the nearly free electron model. Material platforms that mimic each of the two approaches are available for analogue quantum simulation experiments in a scanning tunneling microscope. As the tight-binding model is based on the coupling of atomic orbitals centred at individual sites, structures assembled from closely bound atoms adsorbed on the surface are naturally described using this approach. A second material platform that is reminiscent of the tight-binding model are lattices made of Cl vacancies in NaCl/Cu(111)[84] and in Cl/Cu(100)[42] (Fig. 1a). The electronic state associated with a vacancy can couple to that of a neighbouring vacancy in the same way as orbitals located on close-by atoms can couple.

The nearly free electron model is best suited to describe artificial lattices created by patterning periodic scattering potentials on a 2D electron gas. The most prominent example of a 2D electron gas in STM literature is the Shockley surface state existing on several (111) noble metal surfaces. These surfaces typically host strong quasi-particle interference [85,86] (Fig. 1b). By positioning intentionally introduced surface adsorbates, the 2D electron gas can be scattered and ultimately confined into nearly any geometry defined by the manipulated pattern, as we discuss below. To illustrate the conceptual link between the nearly free electron and the tight-binding models, consider the formation of a ring of metal adatoms on Cu(111) leading to the confinement of the surface state electrons inside the ring (see Box 2)[74,87]. These so-called quantum corrals can either be thought of as particle-in-a-box systems (nearly free-electron model) or as 2D artificial atoms (tight-binding picture). Similarly, coupled quantum corrals and lattices created by patterning a 2D electron gas can be described both using both pictures.

A similar division can be made between two opposing regimes in coupled spin lattices, where the comparable parameter is the coupling strength between individual interacting spins, which is tuned by interatomic separation. The analogue of the tight-binding regime would be the case where individual localized spin states are coupled via nearest-neighbour exchange interactions only, i.e. via direct exchange or superexchange (Fig. 1c). The experimental systems that come closest to this limit are Fe atoms residing at next nearest neighbour sites on metallic surfaces[47,88], as well as the assembly of magnetic adatoms decoupled by an ultrathin insulating film from the conduction electrons in the underlying metal. On $Cu_2N$/Cu(100), transition metal atoms are incorporated into the surface molecular network through which superexchange interactions are mediated[89,90].

The spin-analogue of the nearly free electron model is the situation where the conduction electrons form a continuous spin bath that can mediate correlations between localized spins over a long distance (Fig. 1d). This indirect exchange, or so-called Ruderman-Kittel-Kasuya-Yosida (RKKY) interaction, plays a dominant role when magnetic atoms are placed directly atop a metal surface[1,6,91] with larger interatomic separation. It should be noted that the above division cannot be enforced very strictly in the material systems discussed here. Even for atoms on thin insulators, RKKY interactions are considered to play a sizable role. More prominent still is the role of magnetic anisotropy and Kondo screening, which can result in correlated states across multiple lattice sites if the atoms are placed sufficiently close[92].

*Box 2*: **Atom manipulation – controlling matter one atom at a time**

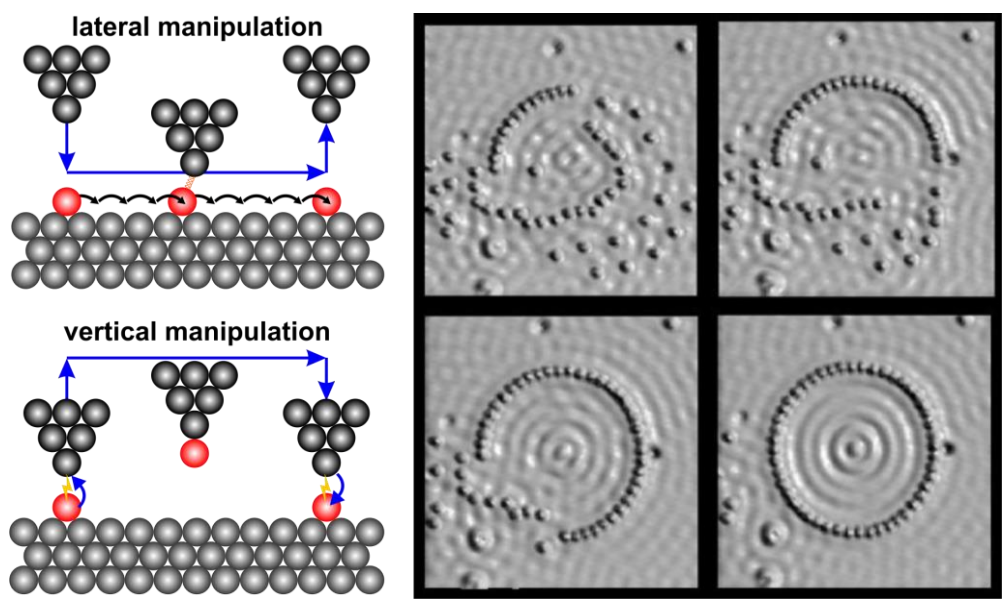

Atomic manipulation is based on the principle that there is a force between the STM tip and the adsorbate when the tip is brought very close. In case of lateral manipulation of atoms or molecules, the tip is first lowered to the adsorbate via reduction of the tunneling resistance, and then moved laterally along the surface. In this regime, the force between the tip and the adsorbate is strong enough to overcome the surface diffusion barrier[51,71,72]. The prototypical case for adatoms on metal surfaces exhibits an attractive tip-adsorbate force, i.e., the adsorbate follows the lateral motion of the tip, which is referred to as pulling or sliding mode[71,73] (see, e.g., the construction of a quantum corral in the figure[74]). In case of repulsive forces (pushing mode), the direction of adsorbate movement is less well controlled, but on a surface with anisotropic diffusion barriers this mode is also very useful[71]. In case of vertical manipulation, the adsorbate is reversibly transferred to the tip and back to the sample using voltage pulses with opposite polarity[75,76]. This mode has the advantage that it allows for adsorbate transfer over very large distances as well as over obstacles such as step edges[76]. Furthermore, it works on decoupling layers where lateral manipulation usually cannot be applied[37]. Atomic manipulation can also be automated, allowing the creation of arbitrary artificial structures with atomic-scale precision[5,50]. A hybrid form of these manipulation techniques is the local application of voltage (or current) pulses, which can induce single-step motion of adsorbates[77]. This strategy is also applied for tip-induced desorption of adsorbates to artificially structure a surface, e.g. local dehydrogenation of H-passivated Si surfaces[11,78]. We note that atom manipulation has also been demonstrated using inelastic tunneling processes[77,79] and atomic force microscopy (AFM)[72,80]. More information about atomic manipulation can be found in several review articles[73,81-83].

**Artificial Electronic Lattices**

In artificially constructed electronic systems it is possible to control the geometry and interaction strength of the artificial atoms in order to study the effects of coupling, topology, strain, periodicity (or lack thereof) and dimensionality. In this section, we will discuss examples of how these degrees of freedom have been used to realize various quantum states of matter.

After the construction of the quantum corral (cf. Box 2) and quantum mirage, which demonstrated the ability to tailor quantum confinement [43,93], one of the first phenomena to be studied bottom-up with atomic manipulation was the evolution of electronic band structure, from the limit of a single atom to extended chains (Fig. 2). One characteristic of quantum confinement in reduced dimensions is the appearance of quantum well states (QWS), where the dispersion of the states is discretized and depends on the confinement potential derived from the lattice constant in the direction of confinement. The appearance of the QWS can be probed as the structure is built. QWS built bottom-up have been investigated in a variety of manipulated chains derived from coupled adatoms on metallic, nearly insulating, and semiconducting surfaces[2,7,94-99]. Moreover, induced atomic-scale defects, which exhibit a degree of charge localization, have also been utilized to create QWS, for example from dangling bonds derived from dehydrogenated silicon atoms[11] and by controllably coupled Cl vacancies[4,14]. A characteristic example is shown in Fig. 2a, where the emergence of QWS can be seen for coupled Cu atoms on a Cu(111) surface. All experiments on metal adatoms on metal surfaces provide qualitatively the same results[2,94-96,99], suggesting that the exact nature of the metal adatoms and metal substrate are not important.

In contrast to metallic surfaces, for coupled adatoms on semiconductors, both direct coupling between adsorbates (Fig. 2b), as well as complex site-specific hybridization of adatoms[98] and substrate-derived states (chains created via self-assembly)[100] have been observed. By assembling chains along different crystallographic directions of a substrate, it is possible to control the distance between the metal adatoms and therefore the hopping parameter between the adatoms[97]. As expected, an increased nearest-neighbor distance results in less coupling and smaller energy level spacing as well as an increased effective electron mass. The conclusions drawn from the work on adatom chains are supported by experiments using defect platforms coupling dangling bonds on a hydrogen-terminated Si(100)[11] (Fig. 2c), and coupling Cl vacancies in Cl/Cu(100) (Fig. 2d).[14] In these systems the hopping parameter is controlled by the spacing of the defects. As expected from basic tight-binding considerations, a shorter lattice spacing results in a stronger hopping parameter and broader bands. The finite 1D chain features standing Bloch waves, indicative of quasiparticle bands. In case of the Cl vacancies on Cl/Cu(100), 2D lattices with different structures were studied as well and it was found that the effective electron mass strongly depends on the lattice geometry.

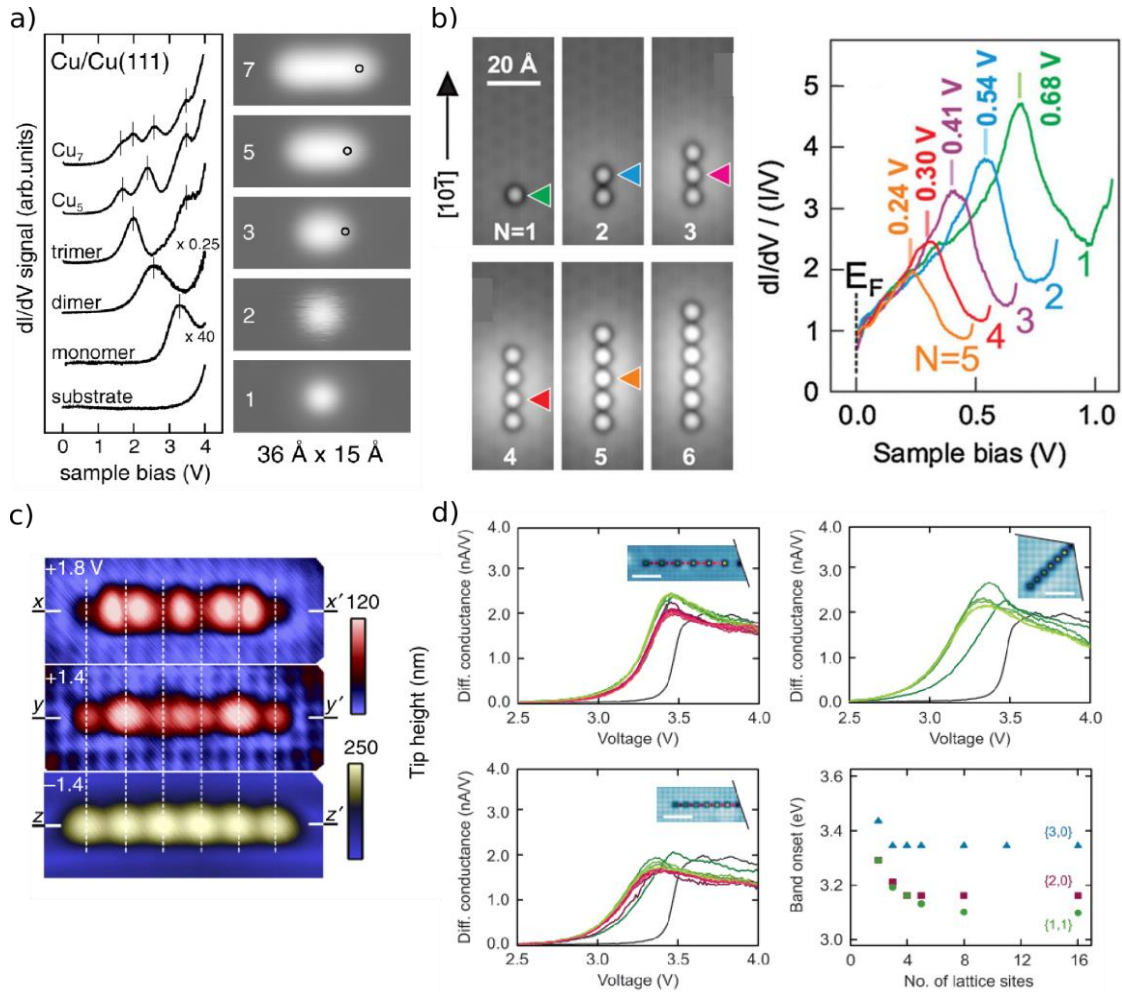

*Fig. 2. Formation of quantum well states from nearest-neighbor interactions of adatoms and vacancies.* (a) Left: differential conductance spectra of chains of Cu atoms on Cu(111) (number of atoms in the chain indicated). Peaks marked by vertical bars indicate the formation of quantized states in the pseudogap of the projected Cu bulk band structure. Right: topographic images of the corresponding adatom chains. Open circles indicate the positions where the dI/dV spectra where acquired. Image taken from ref.[2]. (b) STM images showing the successive assembly of a $In_6$ chain on InAs. The coloured arrows indicate the positions where the differential conductance spectra shown in the right panel were acquired. Image adapted from ref.[7]. (c) STM constant-current images of a chain of five closely spaced dangling bonds on the Si(100):H surface taken at the indicated voltages. Image taken from [11]. The changes in the pattern with energy reflect the different wave functions of the chain. (d) dI/dV spectra taken on vacancy sites and/or chlorine interstices on lattices with 16 vacancy sites for spacing configuration {3,0} (top left), {2,0} (bottom left) and {1,1} (top right). {x,y} indicates the vacancy spacing in x, and y directions, respectively. Locations where spectra were taken are indicated by coloured dots in the insets (blue: Cl atoms, black: vacancies). The bottom right panel shows how the onset of the peak depends on chain length for different configurations. Image taken from ref.[14].

A breakthrough going beyond creating QWS was the realization of artificial or so-called molecular graphene, from a larger ensemble of molecules that can be arranged into 2D lattices to realize more exotic quantum states of matter.[3] The conceptual idea, similar to the quantum corral, is to utilize adsorbates as scattering centres and create a periodic scattering potential pattern. The artificial graphene is the first clear example in which STM manipulation was used to create designer quantum matter (Fig. 3). For the molecular graphene, the 2DEG of the Cu(111) surface state was patterned by arranging CO molecules in a hexagonal lattice on the surface. The formation of the triangular anti-lattice creates the necessary honeycomb geometry in the 2DEG (Fig. 3a, left panel).[3,101,102] The manifestation of molecular graphene is the appearance of a Dirac-like V-shaped LDOS (Fig. 3a, right panel). DFT calculations show that the band structure of the artificial graphene indeed features Dirac dispersion, but they are located at the Γ-point (instead of at the K-point for freestanding graphene).[102] Since the electron density of the surface state is not strongly affected by the presence of adsorbates, the effective number of electrons per lattice site increases when making the unit cell larger. This allows shifting the Fermi energy relative to the Dirac point, i.e., to effectively control the level of doping in the system. Utilizing this, atomically sharp *p-n-p* junctions were created by connecting adjacent molecular graphene lattices with differing lattice constants (Fig. 3b). In addition, it becomes possible to study how strain affects the electronic structure. For graphene, strain that breaks the sublattice symmetry introduces a Gauge field, or so-called pseudo magnetic field, which modifies the electronic structure in the same way as an actual magnetic field, although pseudo magnetic fields can be realized that far exceed the highest possible real fields[103]. In STM-assembled lattices of CO on Cu(111), the effect of triaxial strain on the electronic structure of artificial graphene was studied by simply modifying the arrangement of CO molecules, providing a method to experimentally study the electronic structure of graphene in pseudo magnetic fields up to $B$ = 60 T [3].

Utilizing the aforementioned approach, it is also possible to create designer quantum materials which had not been realized previously. The most prominent example has been the realization of the Lieb lattice, which is characterized by a Dirac dispersion, combined with a flat band at the Dirac point. The Lieb lattice in 2D is a decorated square lattice with three atoms per unit cell. Lieb lattices were created by patterning the 2DEG on the surface of Cu(111) as well as by Cl vacancies in Cl/Cu(100) (Fig. 3c,d)[4,5]. In both cases, differential conductance spectra show a peak due to the flat band. However, these peaks are rather broad, which can be understood by non-negligible next-nearest-neighbor coupling ($t_{NN}$). Interestingly, for the experiments on artificial graphene a much smaller value was found, implying that the magnitude of $t_{NN}$ can be tuned by modifying the pattern of CO molecules.

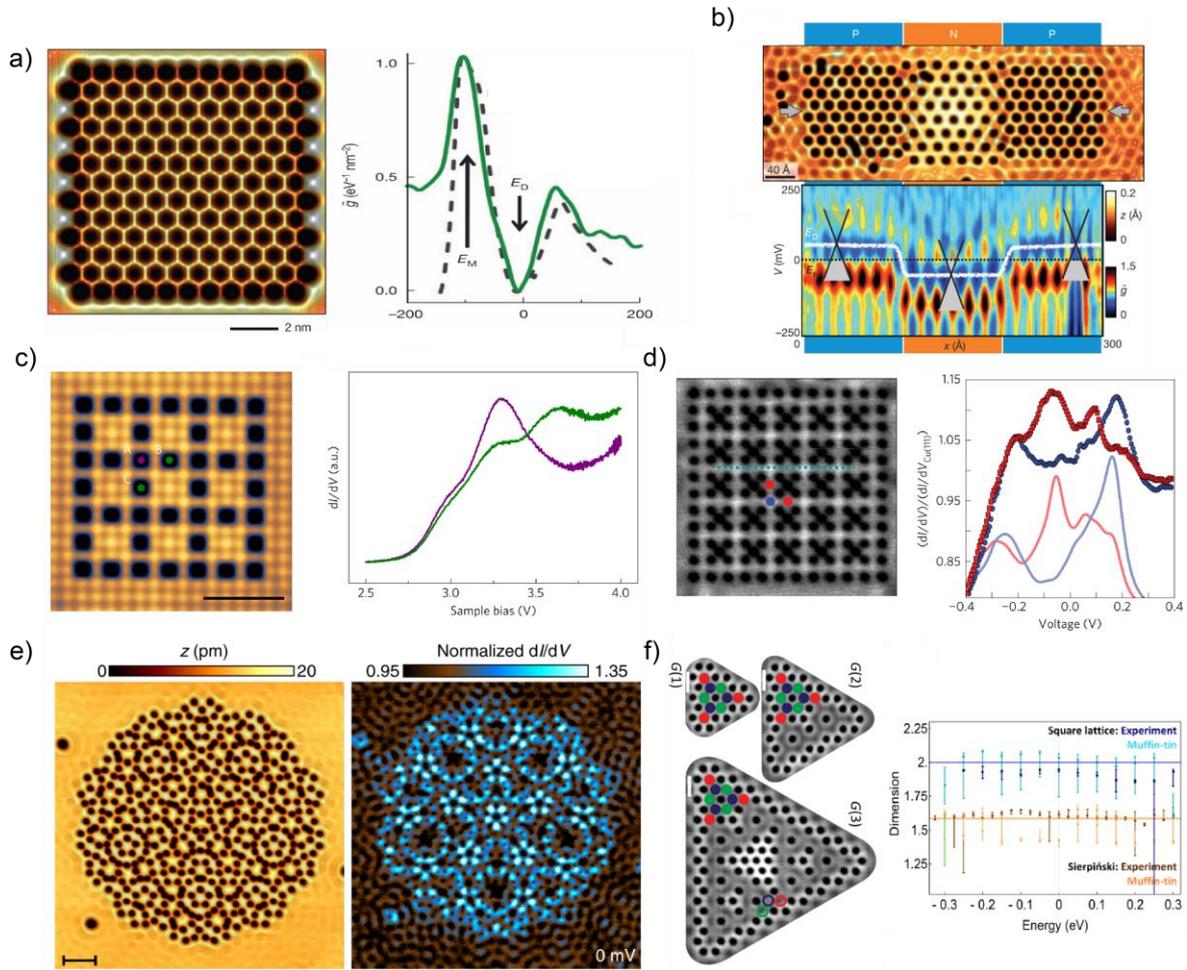

***Fig. 3. Creating complex lattice structures. (a)*** *Left: A hexagonal lattice of CO molecules (black) confines the surface state 2DEG of Cu(111) into its anti-lattice: a honeycomb geometry. Right: dI/dV spectrum (green) of the honeycomb lattice reveals a Dirac-like LDOS. Position of the Dirac cone indicated by $E_D$ The results are corroborated by tight-binding calculations (black dashed line). (b) Top: STM constant-current image of a hexagonal array of CO molecules on Cu(111) with two different spacings between the adsorbates. Bottom: contour plot of differential conductance spectra taken along the line indicated by the grey arrows in the top panel. The position of the Fermi level is indicated by a white line. A larger (smaller) lattice spacing corresponds to the Fermi level being located at a lower (higher) position in the band structure. (c,d) Vacancy lattices (black sites in a) (c)[4] and 2DEGs patterned with CO molecules (black) (d)[5] have been used to create an electronic Lieb lattice. Left panels: STM constant-current images. Right panels: dI/dV spectra taken at positions indicated by the coloured circles in the left panels, (e) Left: STM image of an arrangement of CO molecules (black) that leads to a Penrose tiling for the surface state electrons. Right panel: differential conductance map showing that the local density of states indeed exhibits a Penrose geometry. Images taken from [36] (f) STM images of the first three generations [G(1), G(2) and G(3)] of an electronic Sierpinksi fractal. Different artificial lattice sites are indicated by different colours. Right: plot of the dimension of the electronic states of a G(3) Sierpinksi lattice as determined from experimental differential conductance maps (brown) and tight-binding calculations (orange). For reference, experimental and calculated data on a square lattice is shown as well (dark and light blue, respectively). Images taken from [104].*

Patterned potentials from CO molecules on Cu(111) can also be utilized to probe electronic lattices that lack periodicity: quasicrystals and fractals. Periodicity is one of the fundamental principles underlying our understanding of the electronic structure of materials. By using CO on Cu(111), electrons have been confined to a rhombic Penrose tiling (Fig. 3e)[36]. This quasicrystalline tiling has local order but lacks long-range periodicity. It was found that the electronic wave functions (Fig. 3e, right panel) feature the same symmetries as the geometric structure (left panel): there is rotational symmetry but no translational symmetry. The energy of the states depends on the local vertex structure of the quasicrystal. Fractals are structures that, like quasicrystals, lack periodicity and in many cases also rotational symmetry. They are self-similar on different length scales, known as expanding symmetry, and have a non-integer dimension. Physical fractals are pervasive on the macroscopic scale, but no geometric quantum fractals have been identified. Consequently, very little is known about the behaviour of electrons in non-integer dimensions. The CO on Cu(111) platform was used to create three generations of an electronic Sierpinski fractal.[104] These are shown in the left panel of Fig. 3f. Interestingly, the wave function inherits the dimension of the geometric structure it is confined to (right panels). The self-similarity of the fractal was found in the density of states, in the wave functions, and in reciprocal space.

Artificial electron systems are an ideal platform to test theoretical predictions regarding topology as they provide control over key variables. This is illustrated by experiments on a one-dimensional configuration of Cl vacancies in a Cl monolayer on Cu(100) with alternating weak and strong interactions.[4] Depending on the sequence of interaction strengths, such a chain should feature topologically protected states localized at the ends of the chain with energies in the band gap.[105] Such states were indeed observed.[4]

**Artificial spin lattices**

In addition to the electronic degree of freedom, it is possible to control the spin degree of freedom by utilizing atomic adsorbates which host a magnetic moment and can be manipulated. This enables complex magnetic structures to be constructed, atom-by-atom, and to be probed by spin-sensitive techniques. In this section, we will review various lattices constructed from individual magnetic transition metal atoms, both on metal and thin insulating surfaces. The complexity of these systems ranges from Heisenberg-coupled isotropic spins[37] to strongly anisotropic systems displaying magnetic remanence[45,106] as well as spiralling spin states as a result of Dzyaloshinskii-Moriya interaction[44]. We will also focus on spin arrays that undergo a phase transition due to application of an external magnetic field[6,39].

The artificial spin lattices described here were all probed by either spin-polarized STM (SP-STM)[107] or inelastic electron tunnelling spectroscopy (IETS)[46], or a combination of the two (Fig. 4). These techniques can be applied either in DC mode or in sequences of pulses, giving access to time-dependent phenomena up to about 1 GHz. A commonly used pulse technique is pump-probe spectroscopy[108]. Here, a 'pump' pulse exceeding the voltage threshold for an inelastic process is followed by a 'probe' pulse below the threshold, separated by a time duration $\Delta t$. The spin-polarized contribution due to the probe pulse causes the total current to decay as a function of $\Delta t$, revealing the lifetime of the excited state. A more recent development allows electron paramagnetic resonance (EPR) at even higher frequencies to be performed at the single atom level by means of STM[109], providing access to spin transitions with highly improved energy resolution. In these experiments a high-frequency signal (in the range of 20-30 GHz) is sent to the tip. If the applied frequency matches the Larmor precession of the spin being probed, a Rabi flip of the spin is induced. This, in turn, influences the time-averaged spin-polarized signal, resulting in a sharp peak (or dip) in the current exactly at the resonance frequency. As the focus of this review is on the engineering of spin lattices, we discuss these techniques to the extent that is necessary for the characterization of the interactions of the spins with their environment. A more in-depth discussion of the possibilities enabled in particular by EPR-STM is beyond the scope of this review.

The use of magnetic atoms as building blocks introduces a set of new input parameters for the design of artificial lattices. The most prominent parameters are the magnetic moment, magnetic anisotropy and exchange-type interactions between magnetic moments. As evidenced by examples listed below, each of these can by now be adjusted with varying degrees of controllability. We will focus on four substrates that are commonly used for assembling spin structures: the metal surfaces Pt(111) and Cu(111), and the ultrathin insulators $Cu_2N$ on Cu(100) and MgO on Ag(100) (Fig. 4).

The magnitude of the magnetic moment is determined mainly by the choice of atomic element. As the orbital moment $L$ of 3$d$ transition metal atoms on surfaces is typically quenched, the atomic spin moment $S$ predominantly determines the magnitude of the total magnetic moment found on each atomic site, which we hence refer to as 'spin'. Depending on the level of hybridisation of the moment-bearing orbitals with the substrate conduction electrons, the local spin magnitude can be close to the (half-)integer value of $S$[89] as derived from Hund's rules, reduced in the case of charge transfer[49], or non-half integer in the case of strong hybridization[47]. In the first case, the choice of integer vs. half-integer spin magnitude has profound influence on the expected behaviour of the spin system: due to Kramer's degeneracy, half-integer spin systems are prone to display Kondo screening[110] whereas in integer spin systems such degeneracies are easily broken by single-ion magnetic anisotropy resulting in suppressed or even absent Kondo interactions. In addition, there is the general trend that, the larger the spin, the more profound the influence of magnetic anisotropy. Spin-1/2 systems are even completely immune to magnetic anisotropy.

Magnetic anisotropy is a particularly important design parameter, often described by a typical phenomenological spin Hamiltonian[111]. In situations with low anisotropy, the energy spectrum is typically governed by the Zeeman energy, leading to a ladder-like state distribution. This is the case e.g. for Mn (having $L$ = 0) on $Cu_2N$ [37,68]. Systems with stronger uniaxial anisotropy will exhibit a more barrier-like situation, provided that the anisotropy is of the 'easy-axis' type. This is the case for e.g. Fe (adsorbed on an fcc hollow-site) on Pt(111) [47], Fe on $Cu_2N$ [89] and Co on MgO [112]. Conversely, if the anisotropy is of the 'easy-plane' type, states with small magnetization along the primary anisotropy axis will be favoured, e.g. in Co on $Cu_2N$ [110] and Fe (adsorbed on an hcp hollow-site) on Pt(111) [47].

The third input parameter to be discussed is the magnetic coupling. Depending on the environment, several physical mechanisms may lead to coupling of neighbouring spins. On metal surfaces, the primary interaction is RKKY-like coupling. The oscillatory nature as a function of distance, which is characteristic for RKKY interaction, was measured through magnetisation curves on atoms both on Cu(111) [6] and Pt(111) [1] (Fig. 4a). On Cu₂N, RKKY coupling plays a role too, but in this case there is a competing effect of superexchange coupling through the covalent nitride network making it difficult to disentangle the two effects. Detailed analysis by means of IETS of magnetic dimers built in various orientations revealed that equal separation between the atoms can lead to completely different interaction strengths and even to sign inversion (i.e. antiferromagnetic to ferromagnetic coupling) depending on the crystallographic direction along which the atoms are positioned[9,113] (Fig. 4b). Lastly, on MgO, spin interactions among Fe and Co atoms separated over larger distances (up to several nm) were identified to be of dipolar nature[92,114] (Fig. 4c). The minute µeV-scale shifts in excitation energy due to these interactions were measured via changes in the resonance frequency during an ESR measurement on one of the atoms in the pair.

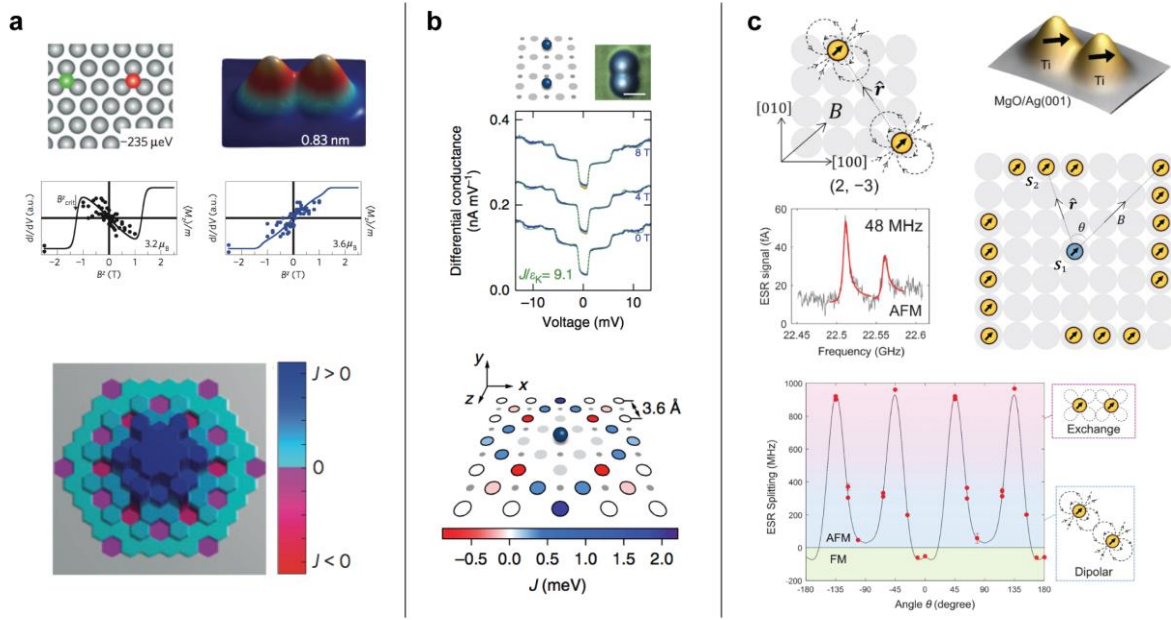

***Fig. 4. Determination of spin couplings on different substrates.*** *(a) Distance-dependent RKKY coupling of atomic Co spins on Pt(111) as measured by SP-STM[1]. Top: Ball model and STM topography of an atom pair. Middle: Magnetisation curves as function of the out-of-plane applied magnetic field on the left (black) and right (blue) atom of the pair; the observed critical field is a measure for the exchange coupling J. Bottom: Calculated values of J for various atom separations. (b) Distance- and orientation-dependent superexchange coupling for two Co atoms on Cu₂N/Cu(100) acquired via spin excitation IETS[9]. Top: Ball model and STM topography of an atom pair. Middle: Differential conductance spectroscopy measurement (blue) and simulation (green) on one of the atoms in the pair. Bottom: Values of J extracted from fitting simulated spectra to measurements for various atom separations. (c) Dipole-dipole coupling of atomic Ti spins on MgO/Ag(001) determined by EPR-STM[12]. Top: Ball model and STM topography of an atom pair. Middle: EPR-STM measurement taken on one atom in the pair (left); the observed splitting of the peak is a measure for J. Overview of various separations investigated, defining the coupling angle θ (right). Bottom: Observed values of J as a function of θ.*

By making use of the design parameters described above and by tuning their relative strengths, it becomes possible to engineer a wide scope of spin lattices that cover different coupling regimes. If for example the spin-spin coupling is chosen to be weak compared to both anisotropy and the external magnetic field, one can realize a situation where neighbouring spins only mildly perturb the local spin of an atom from their isolated state[12,90,92]. When combined with easy-plane anisotropy, which favours states with low magnetization quantum number $m_z$, lattices can be created whose combined low-energy states will be made up exclusively of the low-spin subspace of each of the individual atoms. This technique was used to create an effective spin-1/2 chain from actual spin-3/2 Co atoms on $Cu_2N$[39]. If, on the other hand, this approach is employed in conjunction with easy-axis anisotropy, only the large $m_z$ states enter the low-energy spectrum of the lattice. In such cases, the transverse components of the Heisenberg coupling $J(S_xS_x + S_yS_y) = J/2(S_+S_- + S_-S_+)$ do not to first order couple the states, resulting in a situation that can be effectively described as Ising coupling $JS_zS_z$. Such systems are ideal for probing by means of spin-polarized STM, as the favoured states all have maximum magnetization parallel or anti-parallel to the tip polarization. This technique was used successfully to visualize the transition between various magnetization states in antiferromagnetic and frustrated spin structures which were driven through a phase transition by an external magnetic field[6] (Fig. 5a). This regime was also employed for realizing a logic gate based on antiferromagnetic leads, where the magnetization states of the leads were fixed by large Co islands that served as input gates[38].

When the spin-spin coupling and magnetic anisotropy are chosen to be of comparable strength, a situation arises where the anisotropies of neighbouring atoms reinforce each other. Provided that the anisotropy is easy-axis, this will result in a pair of bistable states that are separated by an energy barrier. The height of this barrier, which determines the switching rate between the two states, depends on the number of spins, their spin magnitudes, their anisotropies and their couplings. Switching rates of 1 Hz and lower were observed for both ferromagnetic[115] and antiferromagnetic[45] Fe chains in $Cu_2N$ (Fig. 5b), and for ferromagnetic arrays on Cu(111)[106] and Pt(111)[116].

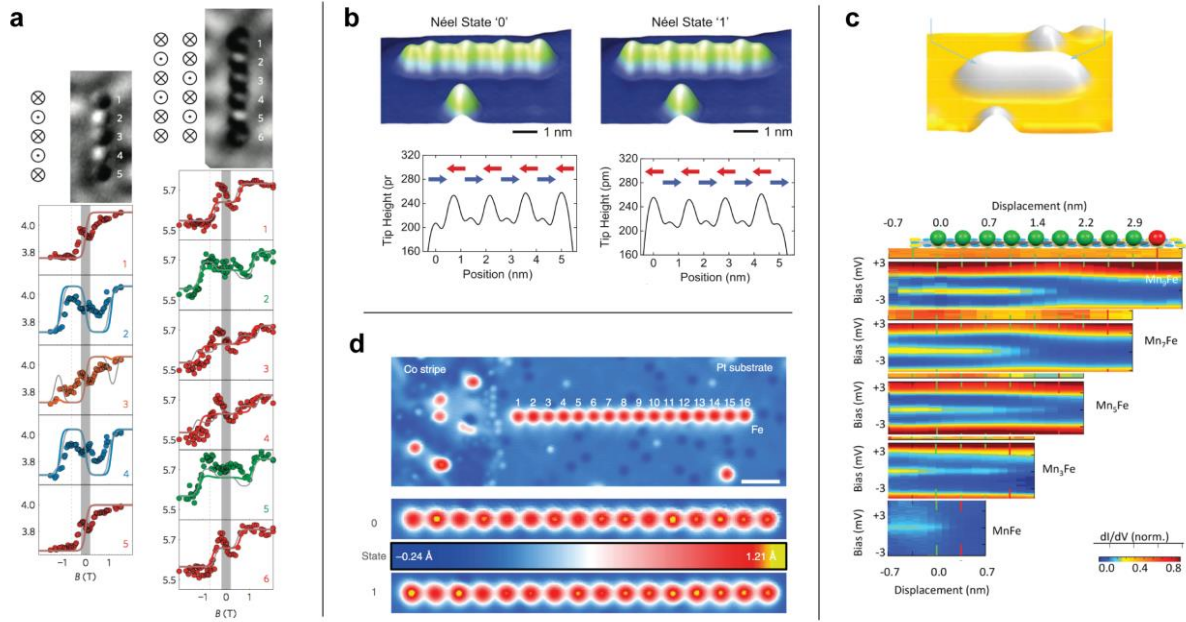

*Fig. 5. Spin chains built in three different coupling regimes.* (a) SP-STM images (top) and corresponding single-atom magnetization curves (bottom) of a 5- and a 6-atom Fe chain on Cu(111) [6]. Colours and arrow diagrams show interpretation of the local magnetisation from the SP-STM contrast. (b) SP-STM images (top) and cross-sections (bottom) of an 8-atom Fe chain on $Cu_2N$/Cu(100) which can be switched between two stable Néel states[45]. Arrows show interpretation of the local magnetisation from the SP-STM contrast. (c) STM constant-current image (top), ball model (middle) and spectroscopic maps (bottom) of a $Mn_xFe$ entangled spin chains on $Cu_2N$/Cu(100) showing the spatial evolution of Kondo screening[117]. (d) STM (top) and SP-STM (bottom) images of a 16-atom Fe chain coupled to a Co stripe on Pt(111), showing non-collinear spin states[44].

The final coupling regime we will discuss is the case where the spin-spin coupling is much stronger than the magnetic anisotropy energy. Spin chains in this regime are often well-described by the giant spin approximation. Spectroscopy in this case typically does not vary from atom to atom but rather describes the excitations of the chain as a whole[37]. Spin chains in this coupling regime were demonstrated to provide a testbed for studying the two-site Kondo effect as a function of separation distance[118]. More recently, spin chains in this coupling regime were locally doped with a single spin of a different magnitude, providing an opportunity for measuring long-range spin correlation and delocalization effects in extended spin states[117] (Fig. 5c).

The coupling mechanisms discussed so far, be it RKKY, superexchange or dipolar interaction, are all collinear couplings and can be expressed in the form of Heisenberg exchange. An additional, non-collinear interaction was found between Fe atoms on Pt(111) (Fig. 5d). This Dzyaloshinskii-Moriya (DM) interaction results from strong spin-orbit coupling with the heavy substrate atoms combined with broken inversion symmetry at the surface. Recently it was shown that the DM interaction can be utilized to engineer spin spirals in sufficiently long chains of spins[44], and provides an interesting additional degree of freedom to create 2D topological magnets, bottom-up, as well as non-collinear magnets on superconductors[119].

**Conclusion and Outlook**

In conclusion, STM can serve as a complete nanoscale lab in which both electronic and magnetic structures can be tailored, atom-by-atom, and subsequently characterized with unprecedented resolution. This enables the construction of novel states of matter, such as Dirac and topologically non-trivial materials which are difficult to engineer[3-5], as well as quasicrystals[36] and electronic fractals[104]. Moreover, complex magnetic order can be tailored bottom-up[6,39,44], and their static and dynamic properties probed with single atom resolution. One important aspect to mention, is the effect of finite size. This was studied systematically for 1D and 2D Cl vacancy lattices, as well as for CO on Cu(111)[5,14]. Surprisingly, finite-size effects are no longer significant for band onsets in Cl-vacancy lattices as small as 8x8 as well as for resonance energies in CO on Cu(111) 5x5 unit cells. Furthermore, similar effects are an important consideration for the development of magnetic order in artificial spin systems, and have been studied in detail in spin chain systems[6,39].

Artificial lattices are promising for studying a variety of complex phenomena, where atomic-scale control of interactions is highly desired toward new theoretical understandings. New potential systems are also emerging, which may serve as new platforms for artificial lattices[92,120]. A natural extension of studying artificial electronic lattices is to look at higher orbital band engineering[121]. This is especially intriguing when spin-orbit coupling is introduced in artificial lattices[122]. Another natural avenue for future exploration is studying topology. Topologically non-trivial states are protected by symmetries of the system. Artificial lattice systems allow for sequentially and selectively breaking all symmetries of the lattice. This platform is therefore ideally suited to experimentally test the protection mechanisms and robustness of topological states. As discussed above, the 1D Su-Schrieffer-Heeger (SSH) model has already been realized by coupling Cl vacancy states[4,123]. For the CO/Cu(111) platform, the interaction strength (hopping parameter) between artificial lattice sites can be tuned by controlling the number and arrangement of the CO molecules that define the boundary between the sites. This approach was used to create a higher order crystalline topological insulator based on the Kagomé lattice[124]. In this lattice, the non-trivial states were found to be localized at the corners of the 2D lattice.

In a similar manner, manipulated spin structures on superconductors is a preferred route toward understanding Majorana zero modes created from engineered magnetic structures on BCS superconductors[125], and a route toward tailoring topological superconductivity[119]. While 1D bottom-up magnetic chains on superconductors are being explored, 2D artificial lattices could also potentially be created[126]. In a similar manner, 2D chiral magnets could be potentially engineered bottom-up[44], to realize isolated topological magnets. For example, it may be conceivable to construct isolated skyrmions with tailored order[127]. Very recent work suggests that it is possible to study heavy-fermion physics by coupling Kondo droplets created by positioning Co atoms on Cu(111)[128].

Moreover, all electronic lattice studies based on STM, to date, are well described within single-particle pictures, and based on tight-binding or muffin-tin approaches[5,101,104,121]. Several other approaches which employ a bottom-up philosophy, such as magic-angle moiré engineering[18,19], cold atoms[8], optical lattices[16,17], and engineered quantum dots[15] have studied many-body ground states driven by electron-electron interactions. Being able to tailor correlated electron effects in artificial lattice systems, atom-by-atom, would be an extremely promising platform for studying emergent ground states, and an essential approach to develop in order to push the artificial lattice platform based on STM methods to compete with other established methods.

Finally, with the development of high temporal resolution techniques[108,109], it will be interesting to study spin and charge dynamics in coupled lattice systems in which hopping can be tailored, as well as magnetic interactions. For example, many-body quantum spin states could potentially be manipulated with microwave or optical methods, by spin resonance[109] or pump-probe methods[108,129], as an outlet for quantum computing. Similarly, the formation of spin frustration as well as glass-like behaviour would be an ideal platform to study the static and dynamic behaviour of spin glasses[130] as well as new directions in stochastic and brain-inspired computing[131].

**Acknowledgements**

AAK would like to acknowledge NWO-VIDI project: "Manipulating the interplay between superconductivity and chiral magnetism at the single-atom level" with project number 680-47-534, and funding from the European Research Council (ERC) under the European Union's Horizon 2020 research and innovation programme (grant agreement No. 818399 'SPINAPSE'). AFO acknowledges support from the European Research Council (ERC Starting Grant 676895 'SPINCAD'). IS acknowledges funding from NWO (grant number 16PR3245-1).